\documentclass[prd,twocolumn,showpacs,amsmath,amssymb]{revtex4-1}
\usepackage[utf8]{inputenc}
\usepackage{amsmath}
\usepackage{latexsym}
\usepackage{amsfonts}
\usepackage{graphicx}
\usepackage{mathrsfs}
\usepackage{CJK}
\usepackage{longtable}
\usepackage{tipa}
\usepackage{multirow}

\usepackage{hyperref}
\hypersetup{
  colorlinks = true,
  urlcolor = blue,
  linkcolor = blue,
  citecolor = green,
  filecolor = magenta,
}

\newcommand{\ud}{\mathrm{d}}
\newcommand{\pd}{\partial}

\begin{document}

\title{Corrections to the gravitational wave phasing}
\author{Shaoqi Hou}
\email{hou.shaoqi@whu.edu.cn}
\affiliation{School of Physics and Technology, Wuhan University, Wuhan, Hubei 430072, China}
\author{Xi-Long Fan}
\email{xilong.fan@whu.edu.cn}
\affiliation{School of Physics and Technology, Wuhan University, Wuhan, Hubei 430072, China}
\author{Zong-Hong Zhu}
\email{zhuzh@whu.edu.cn}
\affiliation{School of Physics and Technology, Wuhan University, Wuhan, Hubei 430072, China}
\affiliation{Department of Astronomy, Beijing Normal University, Beijing 100875,  China}
\date{\today}

\begin{abstract}
  The gravitational wave, traveling a long cosmological distance to reach interferometers, interacts with the (homogeneous and isotropic) cosmological background, so generally speaking, its amplitude and phase are modified in some nontrivial way. 
  As the sensitivity of interferometers is improved, one may detect corrections to the short-wavelength approximation, which naturally includes the information of cosmological evolution. 
  In this work,  the Newman-Penrose variable $\Psi_4$ has been calculated to show that there are two new corrections to the short-wavelength approximation.
  One formally occurs at the first post-Newtonian order but is highly suppressed by the Hubble parameters; the other occurs at the fifth post-Newtonian order, which is due to the variation of the amplitude.
  The first correction contains the evolution of the Universe, but it may not be easily detected.
  The second one indicates that the short-wavelength approximation has to be corrected, when the more accurate waveforms with the higher-order post-Newtonian terms are calculated.
\end{abstract}

\maketitle

\section{Introduction}
\label{sec-intro}

The detection of gravitational waves (GWs) by the LIGO/Virgo Collaborations \cite{Abbott:2016blz,Abbott:2016nmj,Abbott:2017vtc,Abbott:2017oio,TheLIGOScientific:2017qsa,Abbott:2017gyy,LIGOScientific:2018mvr,Abbott:2020uma} confirmed the prediction of general relativity (GR) \cite{Einstein:1916cc,*Einstein:1918btx}, which also marked a new era of GW astronomy and multimessenger astronomy.
The GW is  a probe into the nature of gravity in the dynamical, high-speed regime.
For example, the detection of the polarizations of the GW would exclude either GR or some alternative metric theories of gravity \cite{Will:2014kxa,Hou:2017bqj,Gong:2018cgj,Gong:2018ybk}.
The Kamioka Gravitational Wave Detector (KAGRA) \cite{Somiya:2011np,Aso:2013eba} has now joined the LIGO/Virgo network, and their observations might put a stronger constraint on the GW polarization content.
Moreover, the measurement of the GW speed from GW170817 and GRB 170817A \cite{TheLIGOScientific:2017qsa,Goldstein:2017mmi,Savchenko:2017ffs,Monitor:2017mdv} has already severely constrained some of the alternatives \cite{Baker:2017hug,Creminelli:2017sry,Sakstein:2017xjx,Ezquiaga:2017ekz,Gumrukcuoglu:2017ijh,Gong:2018cgj,Oost:2018tcv,Gong:2018vbo,Casalino:2018tcd,Casalino:2018wnc,Gao:2019liu}.
With the advent of the space-borne and more advanced third-generation detectors such as the Einstein Telescope \cite{Punturo:2010zza}, more GW events can be detected, and our understanding of the nature of gravity, and thereby other phenomena such as cosmology, can be greatly improved.

GWs produced by a compact binary system travel long distances to reach interferometers.
Several interesting propagation effects occur, which have some influence on the amplitude and the phase of the GW.
For the GWs detected by interferometers, their wavelengths $\lambda_\text{gw}$ are  smaller than the Hubble radius $1/H$, so the short-wavelength limit can be safely applied \cite{Caprini:2018mtu}.
When there are no obstacles on the way, the GW amplitude decays, inversely proportional to the luminosity distance \cite{Maggiore:1900zz}.
This is due to the conservation of the number of the gravitons and the expansion of the Universe \cite{Isaacson:1967zz}.
When there are gravitational lenses near the trajectories of the GWs, gravitational lensing takes place \cite{Lawrence1971nc,Lawrence:1971hx,Ohanian:1974ys,Takahashi:2003ix,Liao:2019aqq}.
This causes the magnification of the gravitational amplitudes \cite{gravlens1992}, the rotation of the polarization plane \cite{Ohanian:1974ys,Hou:2019wdg}, and the formation of the beat pattern of the strain \cite{Hou:2019dcm}.

One of the most famous applications of the GW to cosmology is to measure the luminosity distance very accurately; that is, the GW sources are the standard sirens \cite{Schutz:1986gp,Holz:2005df}. 
The first standard siren measurement gave $H_0=70.0^{+12.0}_{-8.0}\text{ km s}^{-1}\text{Mpc}^{-1}$ based on GW170817 \cite{Abbott:2017xzu}, which is consistent with previous measurements, and was improved modestly in a recent joint estimate using the detections made in the first and the second observing runs of LIGO/Virgo \cite{Abbott:2019yzh}. 
Within 5 years, the Hubble constant can be constrained to the 2 \% level with LIGO/Virgo, probably clarifying the Hubble tension and shedding light on dark matter \cite{Chen:2017rfc}.
In order to use the luminosity distance to study some cosmological parameters, one needs to know the redshift of the GW source, which cannot be read off from the waveform obtained based on the short-wavelength approximation.
This is because, in the leading order in the short-wavelength approximation, the gravitational waveform can be put in a form which seems  independent of the source redshift $z$.
Recently, several works took the evolution of the Universe into account, and the waveform carries the information of evolution.
For example, Refs.~\cite{Seto:2001qf,Nishizawa:2011eq} considered the time dependence of $z$, which leads to a nontrivial time evolution of the measured GW frequency.
The resultant waveform acquires a modification -- dephasing -- related to the Hubble parameter and $z$.
This allows one to merely use the GW observation to study cosmology, in principle. 
However, although this effect occurs at $-4$PN order, it is very small, compared to the correction due to the peculiar acceleration of the binary system as discussed in Ref.~\cite{Bonvin:2016qxr}.
So it is not very efficient to make use of this effect to measure $z$.

Besides the effect of the cosmological evolution on $z$, one may consider the other contributions to the dephasing, i.e., corrections to the short-wavelength approximation.
The former correction might occur at the same order (in $\lambda_\text{gw}H$) as the time-varying redshift and was ignored in Refs.~\cite{Seto:2001qf,Nishizawa:2011eq,Bonvin:2016qxr}. 
In the short-wavelength approximation, one writes the metric perturbation in a form with a slowly varying amplitude $A$ and a rapidly oscillating phase $\Phi$, symbolically: $h\sim Ae^{i\Phi}$. 
One usually ignores the slow variation of $A$ to a good approximation, so the GW strain measured by the interferometer is simply proportional to the amplitude $A$.
However, the amplitude varies over space and time, owing to (i) the orbital evolution of the binary system, and (ii) the cosmological expansion.
Taking the changing amplitude into account, one finds out that the measured GW amplitude is modified, as naturally expected.
In addition, one also effectively obtains some new corrections to the phase. 
Since interferometers are very sensitive to the GW phasing, one may want to ignore the amplitude correction but focus on the corrections to the phase.

By directly calculating the Newman-Penrose variable $\Psi_4$, taking the variation of the amplitude into account, one discovers that the variation of the amplitude induces a dephasing in the time domain.
The dephasing consists of two parts. 
The first is due to the  orbital decay of the binary system,  so it becomes stronger as time flies.
The second comes from the cosmological evolution and depends on the Hubble parameters. 
As the GW frequency increases, the second contribution becomes smaller and smaller, as $\lambda_\text{gw}H$ is smaller and smaller. 
In terms of the counting of the post-Newtonian (PN) order, the first effect is at 5PN order, while the second is formally at 1PN order.
In the above calculation, we ignored the peculiar motion of the binary system and the inhomogeneities of the Universe, which were properly considered in Ref.~\cite{Bonvin:2016qxr}. 
Since the 1PN correction discussed in this work also carries a factor of Hubble parameters, it is much smaller than the $-4$PN order correction discovered in Refs.~\cite{Seto:2001qf,Nishizawa:2011eq,Bonvin:2016qxr}, so it can also be ignored.
The 5PN order correction is greater than the 1PN order one, especially in the the frequency bands of ground-based interferometers. 
This correction may be taken into account for more accurate modeling of the GW phase for future advanced detectors.
The computation of the corrections to the short-wavelength approximation in this work is done in a particular coordinate system. 
Several previous works considered the corrections in the full covariant approach \cite{Dolan:2018nzc,Harte:2019tid,Cusin:2019rmt}.

The tidal deformation of neutron stars breaks the degeneracy between the source masses and $z$, also making the electromagnetic counterpart dispensable \cite{Messenger:2011gi,Messenger:2013fya}.
This effect induces corrections to the phase at the 5PN and 6PN orders, but these corrections have large coefficients such that their numerical values are comparable to the 3PN and 3.5PN order terms. 
The method analyzed in Refs.~\cite{Messenger:2011gi,Messenger:2013fya} relies on the knowledge of the neutron star equations of state (EOS), and the predicted accuracy of the redshift measurement is at a few tens percentage level.
The tidal deformability of neutron stars can also be used to obtain the EOS, which requires the accurate waveform at the high PN orders \cite{Kawaguchi:2018gvj}.
The 5PN order correction discovered in this work would inevitably affect the determination of the source redshift and the EOS of neutron stars.

This work is organized in the following way.
In Sec.~\ref{sec-gw-cos}, the propagation of the GW in the Friedmann-Robertson-Walker (FRW) spacetime is derived to obtain the evolution equations for the GW amplitudes in Sec.~\ref{sec-evo}.
Then, the subleading-order GW amplitude is explicitly calculated in Sec.~\ref{sec-bin} for the GW generated by a quasicircular, nonspinning binary star system.
The Weyl tensor is thus computed in order to reveal the corrections due to the varying amplitudes in Sec.~\ref{sec-weyl}.
In Sec.~\ref{sec-fou}, the time-domain waveform obtained in the previous section is Fourier transformed.
The time dependence of the GW frequency is determined in Sec.~\ref{sec-ft}, by mainly reviewing the derivations in Ref.~\cite{Bonvin:2016qxr}.
After that, the stationary-phase approximation is applied to result in the waveform in the frequency domain.
Finally, Sec.~\ref{sec-con} summarizes this work.
In this work, the geometrized units are used with $G=c=1$.

\section{Gravitational Waves in the Cosmological Background}
\label{sec-gw-cos}

In this section, the propagation of the GW in the spatially flat FRW background will be analyzed. 
The GW comes from a binary system, and travels in the Universe, so it interacts with the cosmological background.
The detected GW differs from the initial one in its amplitude and phase.
The evolution of the GW can be studied in the short-wavelength approximation, as the GW wavelengths detectable by interferometers  are much smaller than the Hubble scale.
In the leading order of this approximation, the effects of the cosmological background  include only the decay of the amplitude with the luminosity distance and the redshift of the frequency. 
However, the cosmological evolution, especially the redshift, is completely hidden in the expression for the waveform, once it is written in terms of some locally measurable quantities.
If one goes beyond the leading-order short-wavelength approximation, the cosmological evolution would affect the waveform in an explicit way.  
So in principle it is not necessary to find an electromagnetic counterpart in order to study cosmology.

\subsection{The evolution equations}
\label{sec-evo}

In this subsection, the evolution equations of the GW will be derived up to the first order in the short-wavelength approximation.
The background metric, the FRW metric, is given by \cite{Weinberg:2008zzc}
\begin{equation}
  \label{eq-frw-met}
  \ud s^2=\bar g_{\mu\nu}\ud x^\mu\ud x^\nu=a^2(\eta)[-\ud \eta^2+\delta_{jk}\ud x^j\ud x^k],
\end{equation}
where $a(\eta)$ is the scale factor and $\eta=\int\ud t/a(t)$ is the conformal time. 
The GW is the tensor perturbation $h_{jk}$ to the above metric, and the perturbed metric is 
\begin{equation}
  \label{eq-frw-per-met}
  \ud s'^2=a^2(\eta)[-\ud \eta^2+(\delta_{jk}+h_{jk})\ud x^j\ud x^k].
\end{equation}
It satisfies the following equation of motion \cite{Mukhanov:2005sc}: 
\begin{equation}
  \label{eq-gw-eom}
  h''_{jk}+2\mathcal Hh'_{jk}-\nabla^2h_{jk}=0,
\end{equation}
where the prime means the derivative with respective to $\eta$ and $\mathcal H=a'/a$.
$\nabla^2=\pd_j\pd^j$ is the Laplacian for the 3-Euclidean metric.
The tensor $h_{jk}$ is transverse traceless (TT) \cite{Carroll:2004st}, namely,
\begin{equation}
  \label{eq-tt}
  \pd^kh_{jk}=0,\quad h=\delta^{jk}h_{jk}=0.
\end{equation}
These are actually the TT gauge conditions.
In general, one cannot always make the TT gauge conditions, because of the matter perturbation, unless the perturbed matter stress-energy tensor is traceless.
Here, the TT gauge conditions can be chosen, since there is no matter perturbation. 
In fact, a simple way to check whether the TT gauge conditions can be made is to verify if they are preserved by the equation of motion \eqref{eq-gw-eom}.
It is easy to show that Eq.~\eqref{eq-tt} is preserved by Eq.\eqref{eq-gw-eom}.

Usually, in the cosmology literature, one Fourier transforms $h_{jk}$, as the FRW metric possesses the translation symmetry. 
However, for our purpose, we would like to expand $h_{jk}$ in the following way:
\begin{equation}
  \label{eq-exp-h}
  h_{jk}=\Re\left[ (A_{jk}+\epsilon B_{jk}+\cdots)e^{-i\Phi/\epsilon} \right],
\end{equation}
where $\epsilon\sim \lambda_\text{gw} H\ll1$ for the GW detectable by the interferometers, and $\Re$ means to take the real part.

Substituting this expansion into Eq.~\eqref{eq-gw-eom}, one finds out that, at the leading order [$O(1/\epsilon^2)$], 
\begin{equation}
  \label{eq-null}
  -(l_0)^2+\vec{\boldsymbol l}^2=0,
\end{equation}
where $l_\mu=\pd_\mu\Phi$ is the wave vector and $\vec{\boldsymbol l}$ represents its spatial part.
This relation immediately implies that $l^\nu\nabla_\nu l_\mu=l^\nu\nabla_\mu l_\nu=0$ with $\nabla_\nu$ the covariant derivative of the background metric $\bar g_{\mu\nu}$.
So the GW still propagates at the speed of light along the null geodesic in the cosmological background.
At the next order [$O(1/\epsilon)$], one obtains the evolution equation for $A_{jk}$:
\begin{equation}
  \label{eq-ev-a}
  \tilde l^\mu\pd_\mu A_{jk}+\frac{1}{2}A_{jk}\pd_\mu\tilde l^\mu-a\mathcal Hu^\mu l_\mu A_{jk}=0,
\end{equation}
where $u^\mu=\delta^\mu_0/a$ is the 4-velocity of the comoving fluid in the conformal coordinates.
Finally, the evolution of $B_{jk}$ is given by
\begin{equation}
  \label{eq-ev-b}
  \begin{split}
  \tilde l^\mu\pd_\mu B_{jk}+\frac{1}{2}B_{jk}\pd_\mu\tilde l^\mu-a\mathcal Hu^\mu l_\mu B_{jk}\\
  =-i\left( \frac{1}{2}\pd_\mu\pd^\mu A_{jk}-a\mathcal Hu^\mu\pd_\mu A_{jk} \right),
  \end{split}
\end{equation}
at the order of $O(\epsilon^0)$.
The transverse-traceless conditions \eqref{eq-tt} become 
\begin{gather}
  \tilde l^kA_{jk}=0,\quad \tilde l^kB_{jk}+i\pd^kA_{jk}=0,\label{eq-gc-pd}\\
  \delta^{jk}A_{jk}=\delta^{jk}B_{jk}=0.\label{eq-gc-tl}
\end{gather}
These relations resemble those in Ref.~\cite{Misner:1974qy}.
It seems that these relations imply there are more than two degrees of freedom (DOFs), with $B_{jk}$ representing the extra. 
However, one should realize that $B_{jk}$ actually depends on $A_{jk}$ via Eqs.~\eqref{eq-ev-b} and \eqref{eq-gc-pd}.
So there are still two DOFs given by the transverse-traceless part of $A_{jk}$.

Note that $\tilde l^\mu=\eta^{\mu\nu}l_\nu$, not raised by the background spacetime metric. 
So we put a tilde above the kernel symbol $l$.
In the following, any quantities that can be raised or lowered by $\eta^{\mu\nu}$ and $\eta_{\mu\nu}$ carry the tilde symbol.
It should also be noted that all the evolution equations and the gauge conditions are expressed in terms of the partial derivatives.
So this suggests that it would be easier to do the calculation in the unphysical spacetime, which is flat. 

Before solving the evolution equations for $A_{jk}$ and $B_{jk}$, we would like to first consider the flat spacetime limit.
So, let $a$ approach 1; then $\epsilon$ effectively becomes 0, and thus
\begin{equation}
  \label{eq-exp-h-f}
  h_{jk}=\Re\left( A_{jk}e^{-i\Phi} \right),
\end{equation}
which satisfies the following equation of motion:
\begin{equation}
  \label{eq-gw-eom-f}
  h''_{jk}-\nabla^2h_{jk}=0.
\end{equation}
This is exactly the flat spacetime limit of Eq.~\eqref{eq-gw-eom}.
Here, the prime is actually the partial derivative with respect to the physical time.
Equation~\eqref{eq-gw-eom-f} describes the motion of the \emph{free} GW.
One can easily find out that the wave vector $l_\mu$ is also null and satisfies Eq.~\eqref{eq-null}, so it tangents to the null geodesics.
The evolution equation is Eq.~\eqref{eq-ev-a} with the last term vanishing and $\tilde l^\mu$ being the physical wave vector, i.e.,
\begin{equation}
  \label{eq-ev-a-f}
  \tilde l^\mu\pd_\mu A_{jk}+\frac{1}{2}A_{jk}\pd_\mu\tilde l^\mu=0.
\end{equation} 
The gauge conditions \eqref{eq-tt} become 
\begin{equation}
  \label{eq-tt-a-f}
  l^kA_{jk}=0,\quad \delta^{jk}A_{jk}=0.
\end{equation}
Therefore, $A_{jk}$ is a transverse-traceless tensor, which is just the amplitude of the graviton.
In the case of the cosmological background, the GW is not freely propagating, since the second term in Eq.~\eqref{eq-gw-eom} acts like  friction.
When the GW has a much smaller wavelength than the Hubble radius $1/H$, it barely feels the friction term, as the second term is much smaller than the remaining ones. 
In this case, one can still use Eq.~\eqref{eq-exp-h-f} to describe $h_{jk}$ at the leading order in $\epsilon$.
The evolution equation of $A_{jk}$ is \eqref{eq-ev-a}, which is equivalent to 
\begin{equation}
  \label{eq-ev-a-cov}
  l^\mu\nabla_\mu(a^2A_{jk})+\frac{1}{2}(a^2A_{jk})\nabla_\mu l^\mu=0,
\end{equation}
where $l^\mu=\bar g^{\mu\nu}l_\nu$ is the physical quantity.
This equation resembles Eq.~(6.13) in Ref.~\cite{Isaacson:1967zz}, and the factor $a^2$ appearing in front of $A_{jk}$ in this equation is due to the definition of the metric perturbation, referring to Eq.~\eqref{eq-frw-per-met} where the overall factor $a^2$ should be noted.
The similarity between Eqs.~\eqref{eq-ev-a-f} and \eqref{eq-ev-a-cov} shows that it is the leading term $A_{jk}$ that represents the graviton in the cosmological background \cite{Misner:1974qy}.
At the higher orders, one has to add to $A_{jk}$ the corrections $B_{jk},\cdots$ to take into account the effect of the friction term.
These corrections $B_{jk},\cdots$ actually reflect the fact that the GW is no longer freely propagating, and it interacts with the background.
The interaction, represented by the friction term, disperses the GW further, which is encoded by $B_{jk},\cdots$.

Now, to calculate the evolutions of $A_{jk}$ and $B_{jk}$, it is more convenient to use the spherical coordinate system, because we are interested in GWs coming from a binary star system.
Suppose that the GW emanates from the origin of the coordinate system; then, a suitable tetrad basis $\{\tilde e^\mu_{\hat\alpha}\}=\{\tilde l^\mu,\tilde n^\mu,\tilde x^\mu,\tilde y^\mu\}$ can be chosen to be
\begin{gather}
 \tilde  l^\mu=\gamma_0(1,1,0,0),\label{eq-tet-l}\\
 \tilde  n^\mu=\frac{1}{2\gamma_0}(1,-1,0,0),\label{eq-tet-n}\\
 \tilde  x^\mu=\left( 0,0,\frac{1}{r},0 \right),\label{eq-tet-x}\\
 \tilde  y^\mu=\left(0,0,0,\frac{1}{r\sin\theta}\right),\label{eq-tet-y}
\end{gather}
where $r$ is the comoving distance to the origin where the source is and $\gamma_0$ is a constant, which can be fixed later.
This choice of the tetrad basis is actually suggested by the conformal relation between the FRW metric and the flat one.
Measured by $\eta_{\mu\nu}$, these vectors satisfy $-\tilde l^\mu \tilde n_\mu=\tilde x^\mu \tilde x_\mu=\tilde y^\mu \tilde y_\mu=1$, and the remaining contractions vanish.
In addition,  $\tilde l^\nu\pd_\nu \tilde e^\mu_{\hat\alpha}=0$, which can be verified.
Calculations show that $\pd_\mu \tilde l^\mu=2\gamma_0/r$.
Write $A_{jk}=A^+\tilde e^+_{jk}+A^\times \tilde e^\times_{jk}$ with
\begin{gather}
  \tilde e^+_{jk}=\tilde {\boldsymbol x}_j\tilde {\boldsymbol x}_k-\tilde {\boldsymbol y}_j\tilde {\boldsymbol y}_k,\\
  \tilde e^\times_{jk}=\tilde {\boldsymbol x}_j\tilde {\boldsymbol y}_k+\tilde {\boldsymbol y}_j\tilde {\boldsymbol x}_k,
\end{gather}
where the bold symbols represent the spatial parts.
Here, neither of $A^P$ ($P=+,\times$) has the tilde symbol overhead, since they represent the physical amplitudes \footnote{Note that $a^2h_{jk}$ describes the physical GW in the chosen coordinates. 
The physical tetrad basis contains two normalized spatial vector fields $x^\mu=\tilde x^\mu/a$ and $y^\mu=\tilde y^\mu/a$, so the physical polarization tensors are $e^P_{jk}=a^2\tilde e^P_{jk}$. 
Then, at the leading order in $\epsilon$, $a^2h_{jk}=e^{-i\Phi}\sum_{P=+,\times}A^Pe^P_{jk}=e^{-i\Phi}a^2\sum_{P=+,\times}A^P\tilde e^P_{jk}$.
But at the same order, $h_{jk}=e^{-i\Phi}A_{jk}$, so one obtains $A_{jk}=A^+\tilde e^+_{jk}+A^\times\tilde e^\times_{jk}$.}.
One can check that  $\tilde l^\mu\pd_\mu \tilde e^P_{jk}=0$.
Then, Eq.~\eqref{eq-ev-a} comes 
\begin{equation}
  \label{eq-ev-a-1}
\frac{\gamma_0}{ar}\frac{\ud}{\ud r}(arA^P)=0.
\end{equation}
Here, $r(=\eta)$ is used to parameterize the null geodesic.
This gives the usual law of the decay of the amplitude $A^P\propto1/ar$.

Now, in order to calculate $B_{jk}$, one wants to expand it in the following way: 
\begin{equation}
  \label{eq-exp-b}
  B_{jk}=\sum_{P=+,\times}B^P\tilde e^P_{jk}+B^x\tilde e^x_{jk}+B^y\tilde e^y_{jk}+B^b\tilde e^b_{jk}+B^l\tilde{e}^l_{jk},
\end{equation}
where 
\begin{gather}
  \tilde e^x_{jk}=\tilde {\boldsymbol x}_j\tilde {\boldsymbol l}_k+\tilde {\boldsymbol l}_j\tilde {\boldsymbol x}_k,\\
  \tilde e^y_{jk}=\tilde {\boldsymbol y}_j\tilde {\boldsymbol l}_k+\tilde {\boldsymbol l}_j\tilde {\boldsymbol y}_k,\\ 
  \tilde e^b_{jk}=\tilde{\boldsymbol x}_j\tilde{\boldsymbol x}_k+\tilde{\boldsymbol y}_j\tilde{\boldsymbol y}_k,\\ 
  \tilde e^l_{jk}=\tilde{\boldsymbol l}_j\tilde{\boldsymbol l}_k.
\end{gather}
Because of the traceless condition \eqref{eq-gc-tl}, one knows that $B^b=-\gamma^2_0B^l/2$.
The explicit form of $B_{jk}$ will be determined in the next subsection.

\subsection{The gravitational wave generated by a binary system}
\label{sec-bin}

We are mostly interested in the GW produced by a binary star system and then propagating in the cosmological background. 
In this particular problem, in addition to the expansion parameter $\epsilon$, there is yet a second scale $1/r$ in terms of which the GW is expanded.
Any contributions to the GW of the order of $1/r^n$ with $n\ge2$ decay faster than those of the order of $1/r$, so these higher-order contributions can be ignored. 

In the vicinity of a binary system, it is possible to find a local Lorentz frame where the cosmological evolution barely affects the orbital motion of the stars to a good approximation. 
Suppose the orbit lies in the $xOy$ plane of this frame, and let the stars have masses $m_1$ and $m_2$.
The spins are supposed to be zero.
They move around each other in a circular orbit at the angular frequency $\omega_e$.
Here and below, we use the subscript $e$ to indicate that the very quantity carrying it is measured at the source of the GW.
If the quantity is not associated with $e$, it is measured at some arbitrary location along the GW trajectory.
Then the GW emitted is given by $\bar h_{0\mu}=0$ and 
\begin{gather}\label{eq-hb-bs}
  \bar h_{jk}=\Re\left[\mathcal A_ee^{-i\Phi}\left(
  \begin{array}{ccc}
    -1 & -i & 0 \\
    -i & 1 & 0 \\
    0 & 0 & 0
  \end{array}
  \right)\right],\\
   \mathcal A_e=\frac{4\mathcal M_e}{a_e r_e}(\mathcal M_e\omega_e)^{2/3},
\end{gather}
where  $\mathcal M_e=(m_1m_2)^{3/5}/(m_1+m_2)^{1/5}$ is the chirp mass and $\Phi=-2\int_{t}^{t_c}\omega_e(t')\ud t'+\Phi_c$ is the orbital phase with $\Phi_c$ the fiducial coalescence phase at the fiducial coalescence time $t_c$.
The constant $\gamma_0$ can be fixed to be $\gamma_0=2\omega_ea_e$ such that $l^0$ \footnote{Note that this $l^0$ is the time component of the \emph{physical} wave vector.} is the angular frequency of the GW.
Here, we take only the leading-order amplitude in the Newtonian approximation for simplicity. 
The extension of the current calculation to higher-order post-Newtonian (PN) terms is straightforward but tedious.
At some arbitrary distance $r$, the leading-order wave amplitude is 
\begin{equation}
  \label{eq-wa-r}
  \begin{split}
  \mathcal A(\eta-r,\theta,\phi)&=\frac{4\mathcal M_e}{ar}(\mathcal M_e\omega_e)^{2/3}\\
   &=\frac{4\mathcal M}{d_L}(\pi\mathcal Mf)^{2/3}, 
  \end{split}
\end{equation} 
where $d_L=ar(1+z)$ is the luminosity distance, $\mathcal M=(1+z)\mathcal M_e$ is the redshifted chirp mass, and $f=\omega_e/\pi(1+z)$ is the measured GW frequency.
On the right-hand side, $\mathcal A$ is explicitly written as a function of $u=\eta-r$, $\theta=\iota$ (the inclination angle), and $\phi$.
Along the GW ray, $u$ is a constant, so the decay of the GW is due to the increase of the luminosity distance $d_L$.
At a fixed radial location $r=\text{const}$,  the orbital frequency $\omega_e$ increases, which overcomes the growth in $a$, so the GW amplitude actually increases.

Now, one is ready to calculate $B_{jk}$. 
One should first use the second expression in Eq.~\eqref{eq-gc-pd} to easily obtain 
\begin{gather}
  B^l=B^b=0,\\
  B^x=-i\frac{4\mathcal M_e}{\gamma_0^2a r^2}(\mathcal M_e\omega_e)^{2/3}e^{i2\phi}\sin2\theta,\\
  B^y=\frac{8\mathcal M_e}{\gamma_0^2ar^2}(\mathcal M_e\omega_e)^{2/3}e^{i2\phi}\sin\theta,
\end{gather}
which both decay as $1/r^2$.
So neither of them represents the radiation, and they will be ignored in the following calculation.
Note that, in using Eq.~\eqref{eq-gc-pd}, one does not have to specify any initial conditions, as the second expression in Eq.~\eqref{eq-gc-pd} does not contain any partial time derivative.
Finally, as discussed in the previous subsection, $B_{jk}$ is actually determined by $A_{jk}$. 
Even if $B^x$ and $B^y$ look like the extra polarizations, they are not the independent DOFs, so they do not represent the new polarizations. 
There are still two polarizations, given by the transverse-traceless part of $A_{jk}$, i.e., the graviton. 
The arising of $B^x$ and $B^y$ is due to the interaction between $A_{jk}$ and the cosmological background. 
If the spacetime background is flat, both $B^x$ and $B^y$ vanish identically.

Then, one can use Eq.~\eqref{eq-ev-b} to calculate $B^P$.
After some tedious manipulations and assuming $B^P=0$ initially, one finds out that 
\begin{gather}
  B^+=i\frac{\mathcal M_e}{\gamma_0 ar}(\mathcal M_e\omega_e)^{2/3}e^{i2\phi}E(\eta)(1+\cos^2\theta),\\
  B^\times=-\frac{2\mathcal M_e}{\gamma_0 ar}(\mathcal M_e\omega_e)^{2/3}e^{i2\phi}E(\eta)\cos\theta.
\end{gather}
Here, the function $E(\eta)$ is 
\begin{equation}
  \label{eq-def-e}
  E(\eta)=\mathcal H-\mathcal H_e+\int_{\eta_e}^\eta\mathcal H^2(\eta')\ud\eta'+\frac{4}{r}-\frac{4}{r_e},
\end{equation}
in which, $\frac{4}{r}-\frac{4}{r_e}$ can be dropped.
In terms of $H=\dot a/a$ with the dot denoting the derivative with respect to $t$, $B^P$ can also be written as 
\begin{gather}
  B^+=i\frac{\mathcal M^{2}(\pi\mathcal M f)^{-1/3}}{d_L}e^{i2\phi}E'(t)\frac{1+\cos^2\theta}{2},\\
  B^\times=-\frac{\mathcal M^{2}(\pi\mathcal M f)^{-1/3}}{d_L}e^{i2\phi}E'(t)\cos\theta,
\end{gather}
with 
\begin{equation}
  \label{eq-def-e-1}
  E'(t)=H-\frac{H_e}{1+z}+\frac{1}{1+z}\int_{0}^zH(z')\ud z'.
\end{equation}
The presence of $E(\eta)$ or $E'(t)$ is due to the variation of $A_{jk}$ as well as the coupling of the background curvature with $A_{jk}$; refer to Eq.~\eqref{eq-ev-b}.
Again, $B^+$ and $B^\times$ would vanish if the spacetime background is flat.
In addition, as the frequency $f$ increases, $B^+$ and $B^\times$ decrease, which is consistent with the short-wavelength approximation.
Now, it is ready to determine the modified GW waveform.

\subsection{The Weyl tensor}
\label{sec-weyl}

In the previous subsection, the GW has been calculated, taking into account the corrections to the leading-order short-wavelength approximation. 
However, $\bar h_{jk}$ is not the variable which is directly measured by the interferometer.
For ground-based detectors, the strain $h=-2D^{jk}\int\ud t\int \ud t'R_{tjtk}^\text{gw}$ is measured, where $D^{jk}$ is the detector configuration tensor \cite{Hou:2019wdg}.
In fact, it is the Weyl tensor component $\Psi_4$ that is gauge invariant, and gives the GW strain. 
Since the measurement takes place in the region where cosmological evolution again plays little role, one can calculate $\Psi_4$, assuming the flat spacetime background.
So formally, Eqs.~\eqref{eq-tet-l} - \eqref{eq-tet-y} still define a valid tetrad basis with $\eta$ and $r$ now representing the physical time and distance, respectively.
In addition, $\gamma_0$ should be replaced by $2\pi f$.
To calculate $\Psi_4$, one defines a complex null vector field 
\begin{equation}
  \label{eq-def-mv}
  \begin{split}
  m^\mu=&\frac{1}{\sqrt{2}}(x^\mu-iy^\mu)\\
  =&\frac{1}{\sqrt{2}r}(0,0,1,-i\csc\theta).
  \end{split}
\end{equation}
Note that in this expression, we do not put tildes above the symbols $x$ and $y$, because in this case, the Minkowski metric is the physical one.
$m^\mu$ and its complex conjugate $\bar m^\mu$, together with $l^\mu$ and $n^\mu$, form a Newman-Penrose (NP) tetrad basis \cite{Newman:1961qr,Stephani:2003tm}, which facilitates the calculation of $\Psi_4$.

More specifically, $\Psi_4$ is one component of the Weyl tensor $C_{\mu\nu\rho\sigma}$, i.e., \cite{Stephani:2003tm}
\begin{equation}
  \label{eq-def-p4}
  \Psi_4=C_{\mu\nu\rho\sigma}n^\mu\bar m^\nu n^\rho\bar m^\sigma,
\end{equation}
whose real part $\Re\Psi_4$ represents the $+$ polarization and whose imaginary part $\Im\Psi_4$ corresponds to the $\times$ polarization.
At the leading order [$O(1/\epsilon^2)$], the Riemann tensor for the GW is given by \cite{Harte:2018wni}
\begin{equation}\label{eq-rie-lo}
  {}^{[1]}R^\text{gw}_{\mu\nu\rho\sigma}=-2\Re(l_{[\mu}A_{\nu][\rho}l_{\sigma]}e^{-i\Phi}),
\end{equation}
which is actually the leading-order Weyl tensor ${}^{[1]}C^\text{gw}_{abcd}$, because the spacetime background near the detector is flat approximately.
With this, one can easily calculate the leading-order NP variable $\Psi_4$, i.e.,
\begin{equation}\label{eq-psi4-1}
  \Psi_4^{(1)}=\frac{1}{2}\Re(A^+e^{-i\Phi})+\frac{i}{2}\Re(A^\times e^{-i\Phi}),
\end{equation}
where $A^+$ and $A^\times$ are given by
\begin{equation}
  \label{eq-strains-lo}
  A^+=-\mathcal Ae^{i2\phi}\frac{1+\cos^2\theta}{2},\quad A^\times=-i\mathcal Ae^{i2\phi}\cos\theta,
\end{equation}
respectively.
So, at this order, the Weyl tensor for the GW is given by \cite{Chandrasekhar:1985kt}
\begin{equation}
  C^\text{gw}_{\mu\nu\rho\sigma}=2\Re(\Psi_4^{(1)}l_\mu\wedge m_\nu\wedge l_\rho\wedge m_\sigma),
\end{equation}
with $\wedge$ the wedge product. 
Let the detector carry its own coordinate system $\{t,X^{\hat j}\}$ with the two arms of the interferometer pointing in the $\hat X$ and $\hat Y$ directions, which are two arbitrary unit vectors, perpendicular to each other. 
Then, the electric part of the Weyl tensor is 
\begin{equation}
  \label{eq-cgw}
  \begin{split}
  C^\text{gw}_{0\hat j0\hat k}=&C^\text{gw}_{\mu\nu\rho\sigma}(\pd_t)^\mu(\pd_{\hat j})^\nu(\pd_t)^\rho(\pd_{\hat k})^\sigma\\
  =&2(2\omega)^2\Re(\Psi_4^{(1)} m_{\hat j}m_{\hat k})\\
  =&(2\omega)^2(\Re\Psi_4^{(1)}e^+_{\hat j\hat k}+\Im\Psi_4e^\times_{\hat j\hat k}),
  \end{split}
\end{equation}
where $(\pd_0)^\mu=(\pd/\pd t)^\mu$, $(\pd_{\hat j})^\mu=(\pd/\pd X^{\hat j})^\mu$, and $m_{\hat j}=m_\mu(\pd_{\hat j})^\mu$.
$e^P_{\hat j\hat k}$ are the polarization matrices in the detector frame.
The strain is thus given by \cite{Hou:2019wdg}
\begin{equation}
  \label{eq-st}
  \begin{split}
  h(t)=&-2D^{\hat j\hat k}\int\ud t'\int\ud t''C^\text{gw}_{0\hat j0\hat k}\\
  =&2\left(\Re\Psi_4^{(1)}F^++\Im\Psi_4^{(1)}F^\times\right),
  \end{split}
\end{equation}
where $D^{\hat j\hat k}=\left(\hat X^{\hat j}\hat X^{\hat k}-\hat Y^{\hat j}\hat Y^{\hat k}\right)/2$ is the detector configuration matrix and $F^+$ and $F^\times$ are the antenna pattern functions for the + and the $\times$ polarizations, respectively \cite{Poisson2014}:
\begin{equation}
  F^P=D^{\hat j\hat k}e^P_{\hat j\hat k}.
\end{equation}
One recognizes $h^+=A^+e^{-i\Phi}$ and $h^\times=A^\times e^{-i\Phi}$ very easily from Eq.~\eqref{eq-psi4-1}.
Note that the factor $2\omega^2$ in the last line of Eq.~\eqref{eq-cgw} drops out in Eq.~\eqref{eq-st}.
This is because both terms $\Re\Psi_4^{(1)}$ and $\Im\Psi_4^{(1)}$ have the phases $2\omega t+\phi_0^{+/\times}$ with $\phi_0^{+/\times}$ the initial phases for the $+$ and $\times$ polarizations, and each time integral absorbs one factor of $2\omega$.

At the next leading order [$O(1/\epsilon)$], the Weyl tensor is given by a somewhat more complicated expression,
\begin{equation}
  \label{eq-w-2}
  \begin{split}
    {}^{[2]}C^\text{gw}_{\mu\nu\rho\sigma}=&-2\Re(l_{[\mu}B_{\nu][\rho}l_{\sigma]}e^{-i\Phi})\\ 
    &+\Im[(l_{[\rho}\nabla_{|\nu|}A_{\sigma]\mu}
    +l_\nu\nabla_{[\rho}A_{\sigma]\mu}+A_{\mu[\sigma}\nabla_{\rho]}l_\nu\\
    &+\langle\mu\leftrightarrow\nu,\rho\leftrightarrow\sigma\rangle)e^{-i\Phi}],
  \end{split}
\end{equation}
where the symbol $\langle\mu\leftrightarrow\nu,\rho\leftrightarrow\sigma\rangle$ means the terms obtained by interchanging $\mu$ and $\nu$, and, at the same time, $\rho$ and $\sigma$ of the terms in the second line. 
Then, one has  \cite{Martin-Garcia:2007bqa,MartinGarcia:2008qz,xperm2008,Brizuela:2008ra,xact}
\begin{equation}\label{eq-psi4-2}
  \begin{split}
  \Psi_4^{(2)}=&\frac{1}{2}\Re\left(B^+e^{-i\Phi}\right)+\frac{i}{2}\Re\left(B^\times e^{-i\Phi}\right)\\
  &+\frac{2\bar\mu+5\gamma-3\bar\gamma}{2}\left[\Im\left(A^+e^{-i\Phi}\right)+i\Im\left(A^\times e^{-i\Phi}\right)\right]\\
  &+n^\mu\left[\Im\left(e^{-i\Phi}\nabla_\mu A^+\right)+i\Im\left(e^{-i\Phi}\nabla_\mu A^\times\right)\right],
  \end{split}
\end{equation}
where $\sigma=-m^\mu m^\nu\nabla_\nu l_\mu$, $\gamma=(l^\mu n^\nu\nabla_\nu n_\mu-m^\mu n^\nu\nabla_\nu\bar m_\mu)/2$, and $\mu=\bar m^\mu m^\nu\nabla_\nu n_\mu$ are the spin coefficients of the flat metric in the spherical coordinate system \cite{Stephani:2003tm}, $\Im$ stands for the imaginary part, and the overhead bar indicates the complex conjugation.
The straightforward calculation shows that these spin coefficients decay as $1/r^5$, so the second line in Eq.~\eqref{eq-psi4-2} can be ignored, as it does not represent the radiation. 
Then one obtains 
\begin{equation}
  \label{eq-psi4-2-1}
  \begin{split}
    \Psi_4^{(2)}=&\frac{1}{2}\Re\left[-i\left(\frac{E'}{4\pi f}+\Upsilon \right)A^+e^{-i\Phi}\right]\\
                &+\frac{i}{2}\Re\left[-i\left(\frac{E'}{4\pi f}+\Upsilon \right)A^\times e^{-i\Phi}\right].
  \end{split}
\end{equation}
Here, the symbol $\Upsilon$ is
\begin{equation}
  \label{eq-def-Y}
  \Upsilon=\frac{32}{5}(\pi \mathcal Mf)^{5/3}-\frac{H}{4\pi f},
\end{equation}
which comes from the third line in Eq.~\eqref{eq-psi4-2}.
In doing the calculation, we used the following evolution of the GW angular frequency \cite{Maggiore:1900zz}:
\begin{equation}
  \label{eq-ev-f}
  \frac{\ud\omega}{\ud t}=\frac{192}{5}\mathcal M^{5/3}\left( \frac{\omega}{2} \right)^{11/3}.
\end{equation}
In the end, since $B^x$ and $B^y$ are nonvanishing, some of the Weyl components $\Psi_n$ with $n\ne0$ might be nonzero. 
However, since $B^x$ and $B^y$ scale as $1/r^2$, these $\Psi_n$ will also be of the higher order in $1/r$ and are, thus, ignored in the following discussion.

Finally, $\Psi_4$ is given by
\begin{equation}
  \label{eq-psi4-t}
  \begin{split}
  \Psi_4=&\Psi_4^{(1)}+\Psi_4^{(2)}\\
  =&\frac{1}{2}\Re\left\{\left[1-i\left(\frac{E'}{4\pi f}+\Upsilon \right)\right]A^+e^{-i\Phi}\right\}\\
  &+\frac{i}{2}\Re\left\{\left[1-i\left(\frac{E'}{4\pi f}+\Upsilon\right) \right]A^\times e^{-i\Phi}\right\},
  \end{split}
\end{equation}
up to the order of $O(\epsilon)$.
By comparing this expression with Eq.~\eqref{eq-psi4-1}, one can redefine the amplitudes $A^P$ and the phase $\Phi$ such that Eq.~\eqref{eq-psi4-t} takes the same form as Eq.~\eqref{eq-psi4-1}. 
Therefore, the modified amplitudes and phase are
\begin{gather}
  A'^P=\sqrt{1+\left(\frac{E'}{4\pi f}+\Upsilon\right)^2}A^P,\label{eq-hn-am}\\
  \Phi'=\Phi+\arctan\left(\frac{E'}{4\pi f}+\Upsilon\right).\label{eq-hn-ph}
\end{gather}
Since both $E'/4\pi f$ and $\Upsilon$ are small quantities, one may ignore the correction to the amplitudes.
However, because of the sensitivity of interferometers to the phase, one wants to keep the correction to the phase and approximates it as 
\begin{equation}
  \label{eq-dep}
  \begin{split}
  \Phi'\approx&\Phi+\frac{E'}{4\pi f}+\Upsilon\\
   =& \Phi-\frac{1}{4\pi f}\left[H+\frac{H_e}{1+z}-\frac{1}{1+z}\int_0^zH(z')\ud z'\right]\\
   &+\frac{32}{5}(\pi\mathcal Mf)^{5/3}.
  \end{split}
\end{equation}
Therefore, the correction to the phase consists of two parts, formally. 
One is given by the terms in the square brackets, which decreases with the frequency $f$ and conforms to the short-wavelength approximation.
The second part is the last term.
It is a monotonically increasing function of the frequency $f$, which comes from the last line of Eq.~\eqref{eq-psi4-2}.
Note that due to Eq.~\eqref{eq-tet-n}, $n^\mu$ is inversely proportional to $f$ as $\gamma\propto f$.
So, if the terms in the square brackets of that line were proportional to $f^\alpha$ with $\alpha<1$, then that last line would have behavior compatible with the short-wavelength approximation. 
However, $A_{jk}$ increases with $f$ at a fast enough rate [refer to Eq.~\eqref{eq-ev-f}] so that the last line of Eq.~\eqref{eq-psi4-2} actually increases with $f$.
The orbital decay of the binary system causes the increase in $A_{jk}$, which leads to the monotonically increasing contribution to the dephasing. 

In fact, there is yet a hidden correction to the phasing, which is the modified time dependence of the frequency $f$ as nicely derived in Ref.~\cite{Bonvin:2016qxr}.
This hidden correction will be briefly reviewed in the next section, and then, one finds out that it explicitly shows up in the Fourier transform of the phase.

\section{The Fourier transform}
\label{sec-fou}

In this section, the Fourier transform of the GW will be found, which is used for matched filtering \cite{Droz:1999qx}.
First, we will review how the measured GW frequency  varies with the time, based on the discussion in Ref.~\cite{Bonvin:2016qxr}. 
In this review, we ignore the peculiar motion of the binary system and the inhomogeneities of the Universe.
Second, we will derive the Fourier transform of the waveform represented by Eqs.~\eqref{eq-hn-am} and \eqref{eq-dep}. 

\subsection{The time dependence of the frequency}
\label{sec-ft}

When taking the Fourier transformation of the time-domain waveform, it is very important to know how the GW frequency $f$ varies with time $t$.
In the local Lorentz frame of the binary system, the GW frequency $f_e$ follows
\begin{equation}
  \label{eq-fe-dt}
  \frac{\ud f_e}{\ud t_e}=\frac{96}{5}\frac{f_e}{\mathcal M_e}(\pi\mathcal M_ef_e)^{8/3},
\end{equation}
at the leading post-Newtonian order.
From this, one can calculate the phase $\Phi$ in terms of $f=f_e/(1+z)$,
\begin{equation}
  \label{eq-p-f}
  \Phi=-\frac{(\pi\mathcal Mf)^{-5/3}}{16}+\Phi_c.
\end{equation}
Since $\ud t=(1+z)\ud t_e$, one gets the equation satisfied by $f$, 
\begin{equation}
  \label{eq-f-dt}
  \frac{\ud}{\ud t}[(1+z)f]=\frac{96}{5}\frac{(1+z)f}{\mathcal M}(\pi\mathcal Mf)^{8/3},
\end{equation}
from Eq.~\eqref{eq-fe-dt}.
In the usual approach, one sets $z$ to be a constant, so one can take it out of the time derivative on the left-hand side of the above expression, such that $f$ evolves with $t$ exactly the same way as $f_e$ does with $t_e$, with appropriate rescaling of some quantities by powers of $(1+z)$.
This is a good approximation because of the much faster orbital evolution than the cosmological expansion. 
However, if one wants to calculate the corrections to this approximation, the result is different. 
Directly integrating Eq.~\eqref{eq-f-dt} leads to 
\begin{equation}
  \label{eq-ft}
  [(1+z)f]^{-8/3}=\frac{256}{5}\frac{(\pi\mathcal M_e)^{8/3}}{\mathcal M_e}\int_t^{t_c}\frac{a(t'_e)}{a(t')}\ud t'.
\end{equation}
During the lifetime of the binary system, the scale factor $a$ changes by a small amount, so one can approximate it in the following way:
\begin{gather}
  a(t'_e)\approx a(t_e)+(t'_e-t_e)\frac{\ud a(t_e)}{\ud t_e},\\
  a(t')\approx a(t)+(t'-t)\frac{\ud a(t)}{\ud t}.
\end{gather}
Substituting these into Eq.~\eqref{eq-ft} and ignoring higher-order terms in $(t'-t)$ and $(t'_e-t_e)$, one obtains
\begin{equation}
  \label{eq-f-t}
  f=\frac{1}{\pi\mathcal M}\left( \frac{256\tau}{5\mathcal M} \right)^{-3/8}\left[ 1+\frac{3}{8}X(z)\tau \right],
\end{equation}
where $\tau=t_c-t$ and 
\begin{equation}
  \label{eq-xz}
  X(z)=\frac{1}{2}\left( H-\frac{H_e}{1+z} \right).
\end{equation}
The second term in the square brackets in Eq.~\eqref{eq-f-t} represents the effect of the cosmological evolution on the time dependence of $f$.

In the above discussion, one does not consider the dephasing given in Eq.~\eqref{eq-dep}.
Once this dephasing is taken into account, the effective GW frequency is 
\begin{widetext}
 \begin{equation}
  \label{eq-ef-f}
  \begin{split}
  f'=&\frac{\ud\Phi'}{\ud t}\\
  =&f+\frac{3}{8f}\left[ \frac{256}{5\mathcal M}(\pi\mathcal Mf)^{8/3}-2X\right]\left[ \frac{5H}{4\pi}-\frac{H_e}{1+z}+\frac{1}{1+z}\int_0^zH(z')\ud z' \right]-\frac{3968}{5}(1+z)(\pi\mathcal Mf)^{5/3}X\\
  &-\frac{1}{f}\left[ \frac{5}{4\pi}\frac{\ud H}{\ud t}-\frac{1}{(1+z)^2}\frac{\ud H_e}{\ud t_e}+2X\left( H+\frac{H_e}{1+z} \right)-\frac{2X}{1+z}\int_0^zH(z')\ud z' \right]+\frac{2048}{5\mathcal M}(1+z)(\pi\mathcal Mf)^{13/3}.
  \end{split}
\end{equation} 
\end{widetext}
To obtain this expression, one uses the following useful results:
\begin{gather}
  \frac{\ud z}{\ud t}
  =2(1+z)X,\\
  \frac{\ud X}{\ud t}=\frac{1}{2}\left[ \frac{\ud H}{\ud t}-\frac{1}{(1+z)^2}\frac{\ud H_e}{\ud t_e}+\frac{2H_eX}{1+z} \right],\\
  \frac{\ud f}{\ud t}=\frac{3f}{4}\left[ \frac{128}{5\mathcal M}(\pi\mathcal Mf)^{8/3}-X \right],
\end{gather}
which can be easily obtained via straight forward but tedious calculations.
Therefore, inspecting Eq.~\eqref{eq-ef-f} shows that the corrections to the effective GW frequency $f'$ are of the higher order, and one simply approximates $f'\approx f$ in the following.

\subsection{The stationary-phase approximation}
\label{sec-spa}

The Fourier transformation of the time-domain GW can be done using the stationary-phase approximation, i.e., the saddle-point approximation \cite{Droz:1999qx,Yunes:2009yz}.
This is due to the much faster changing phase $\Phi'$ relative to the amplitudes $A'^P$.

Let us start with the Fourier transformation of a generic signal $s(t)=\Re\{\hat A(t)e^{-i\hat\Phi(t)}\}$, whose phase $\hat\Phi(t)$ varies with time $t$ rapidly. 
The Fourier transformation is 
\begin{equation}
  \label{eq-ft-s}
  \begin{split}
    \tilde s(F)=\int_{-\infty}^\infty\hat A(t)e^{i(2\pi Ft-\hat\Phi(t))}\ud t.
  \end{split}
\end{equation}
The saddle point $t_s$ is determined by the extreme of the exponent in the integrand, $2\pi F-\ud\hat\Phi(t_s)/\ud t=0$, i.e., $f_s\equiv\ud\Phi(t_s)/\ud t=F$.
One can then carry out the above integration in the frequency domain, i.e., replacing $\ud t=\ud f/\dot f$.
Also, one expands the exponent around the saddle point $f_s$, so 
\begin{equation}
  \label{eq-ft-s-f}
  \begin{split}
    \tilde s(f_s)=&\int_{-\infty}^\infty\frac{\hat A(t)}{\dot f}\times \\
    &\exp\left\{i\left[2\pi f_st_s-\hat\Phi_s-\frac{\dot{f}_s^2}{2}(f-f_s)^2\right]\right\}\ud f,
  \end{split}
\end{equation}
where $\hat\Phi_s\equiv \hat\Phi(t_s)$, and $t$  in the above expression is taken to be an implicit function of $f$.
Simply carrying out the integration gives 
\begin{equation}
  \label{eq-ft-s-1}
  \begin{split}
    \tilde s(F)=\frac{\hat A(t_s)}{\sqrt{\dot f_s}}\exp\left[i\left(2\pi f_st_s-\hat\Phi_s-\frac{\pi}{4}\right)\right],
  \end{split}
\end{equation}
from which, one can read off the phase of the Fourier transformed signal,
\begin{equation}
  \label{eq-ph-f}
  \tilde{\hat\Phi}=2\pi f_st_s-\hat\Phi_s-\frac{\pi}{4},
\end{equation}
with $t_s$ an implicit function of $f_s$.
So, if one knows the dependence of the frequency $f_s$ on $t_s$, one can invert it to obtain $t_s=t_s(f_s)$.

\begin{figure}[h]
  \centering
  \includegraphics[width=0.4\textwidth]{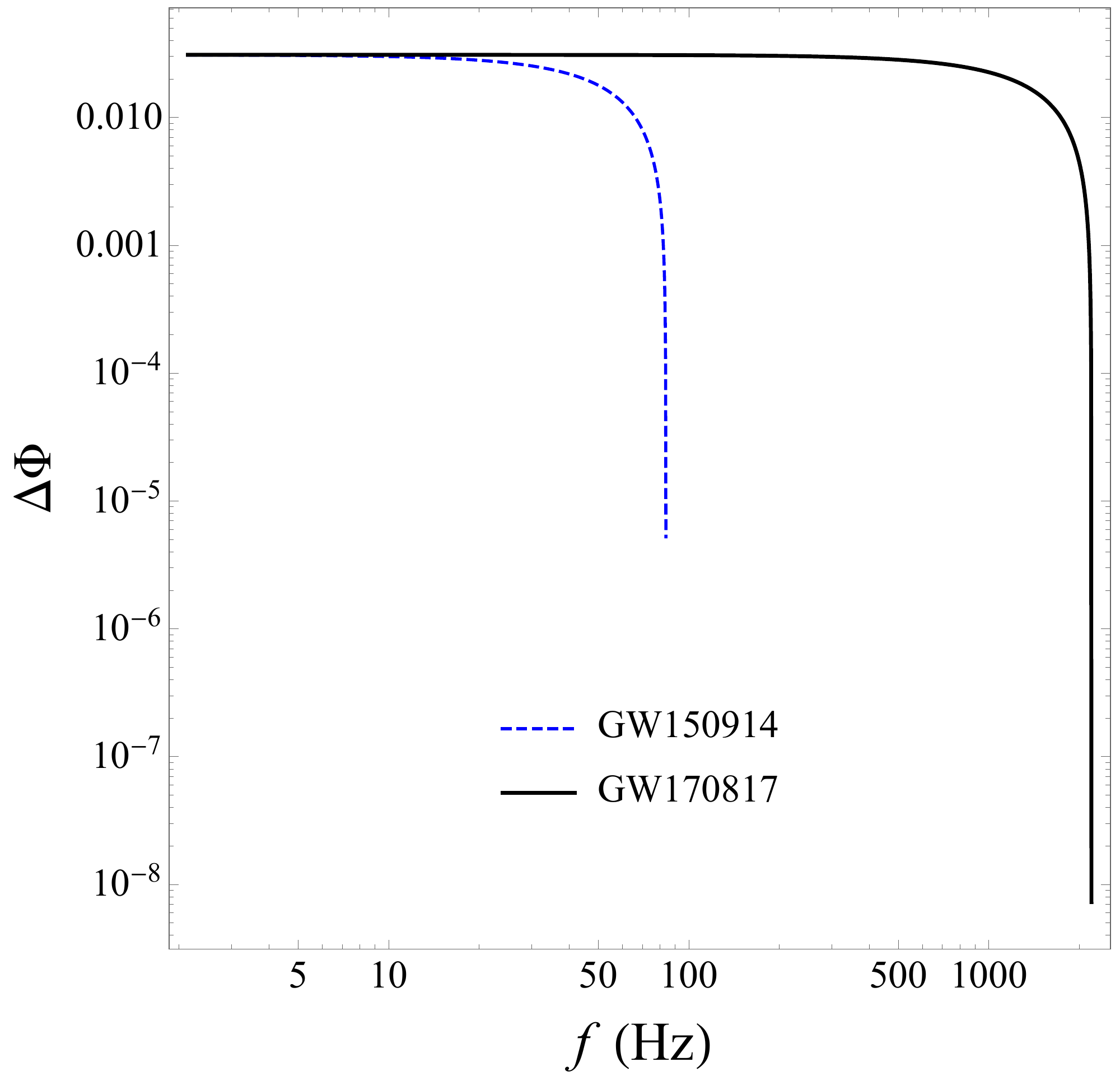}
  \includegraphics[width=0.4\textwidth]{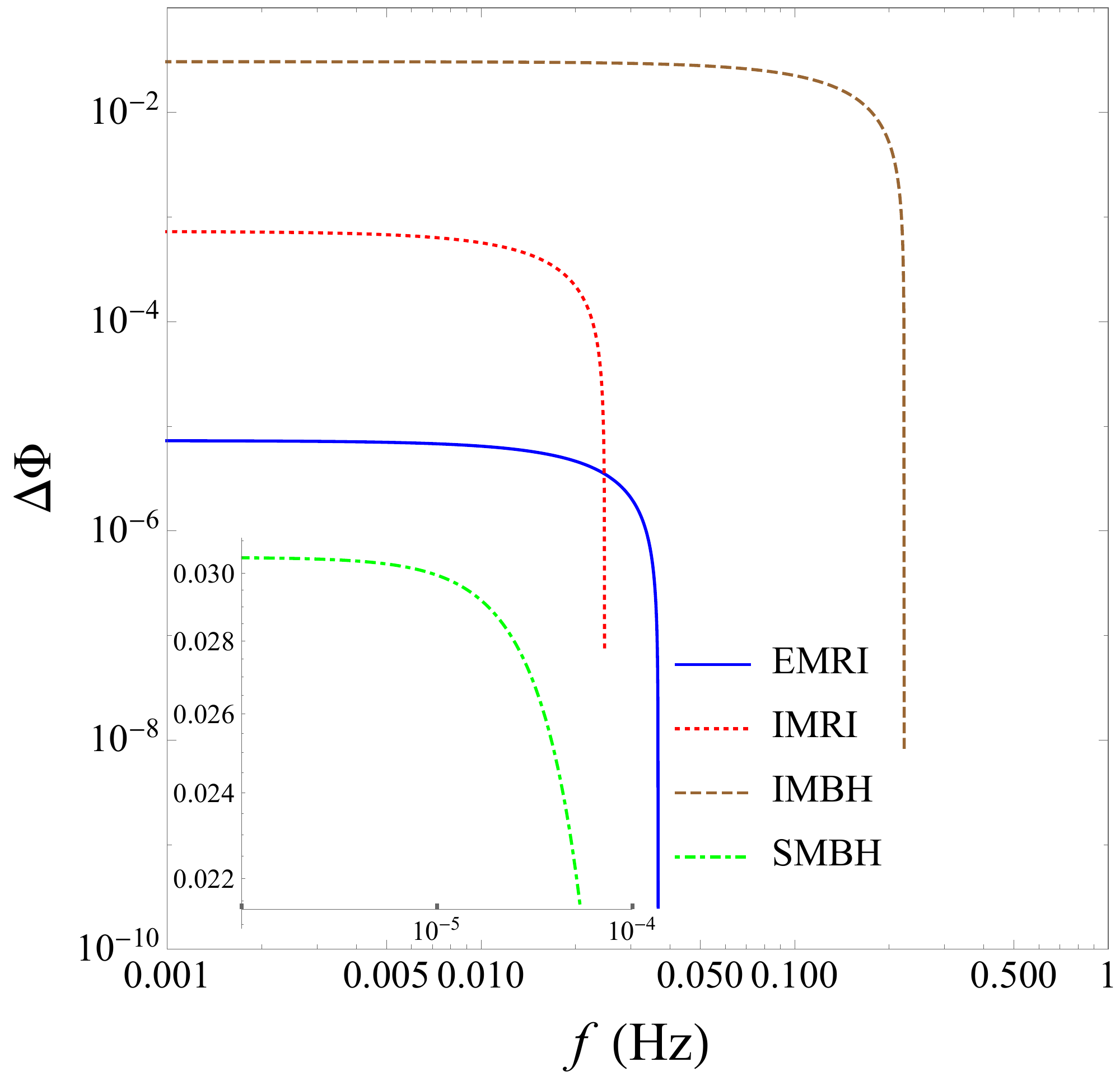}
  \caption{The dephasing $\Delta\Phi$ to be accumulated due to the 5PN correction as a function of the initial observation frequency $f$.
  The upper panel shows $\Delta\Phi$ for GW150914 and GW170817 \cite{LIGOScientific:2018mvr}.
  The lower panel shows $\Delta\Phi$ for sources defined in Table~\ref{tab-pars}.}
  \label{fig-dep}
\end{figure}

Applying this general prescription to the GW strain \eqref{eq-hn-am} and \eqref{eq-dep}, one finds out that the Fourier transformed GW has the following phase: 
\begin{equation}
  \label{eq-gw-ph}
  \begin{split}
  \tilde \Phi'=&2\pi ft_c-\Phi_c-\frac{\pi}{4}+\frac{3}{128}(\pi\mathcal Mf)^{-5/3}\\
  &-\frac{25\mathcal M}{32\,768}(\pi\mathcal Mf)^{-13/3}X(z)\\
  &-\frac{1}{4\pi f}\left[ H+\frac{H_e}{1+z}-\frac{1}{1+z}\int_0^zH(z')\ud z' \right]\\
  &+\frac{32}{5}(\pi\mathcal Mf)^{5/3}.
  \end{split}
\end{equation}
The first line is the usual leading-order phase presented in Ref.~\cite{Buonanno:2009zt}.
The second line was already given in Ref.~\cite{Bonvin:2016qxr}, which is due to the modified time dependence of the measured GW frequency $f$, as a result of the cosmological evolution of $z$.
This term is formally at $-4$PN order, because $(\pi\mathcal Mf)^{1/3}$ is nothing but the characteristic (linear) velocity $v$ of the binary system, and the last term in the first line is at 0PN order \cite{Buonanno:2009zt}. 
The ratio between this term and the last one in the first line is proportional to $(\pi \mathcal Mf)^{-8/3}=v^{-8}$, so this term is at $-4$PN order.
The last two lines are the corrections to the leading-order short-wavelength approximation.
The third one is due to the cosmological evolution, and formally, is at 1PN order. 
As explained in the penultimate paragraph in Sec.~\ref{sec-weyl}, the fourth line comes from the orbital decay of the binary system and is at 5PN order.
By setting $f$ to be the GW frequency corresponding to the innermost stable orbit \cite{Bonvin:2016qxr}
\begin{equation}
  \label{eq-fisco}
  f_\text{isco}=8.80(1+1.25\eta+1.08\eta^2)\left[ \frac{M_\odot}{(1+z)(m_1+m_2)} \right]\text{kHz},
\end{equation}
with $\eta=m_1m_2/(m_1+m_2)^2$,  the 5PN order correction contributes 
\begin{equation}
  \Delta\Phi_\text{5PN}^\text{max}\approx0.099
\end{equation}
at most up to the frequency $f_\text{isco}$.
Figure~\ref{fig-dep} shows the dephasing $\Delta\Phi$, to be accumulated from some initial observation frequency $f$ to $f_\text{isco}$, due to the 5PN correction for different types of sources defined in Table~\ref{tab-pars}.
These sources include the binary star systems observed in GW150914 and GW170817 for ground-based interferometers \cite{LIGOScientific:2018mvr} and the sources for space-borne detectors: the extreme mass-ratio inspiral (EMRI), the intermediate mass-ratio inspiral (IMRI), the intermediate mass black hole binary (IMBH) and the supermassive black hole binary (SMBH) \cite{Chamberlain:2017fjl}. 
From Fig.~\ref{fig-dep}, one finds out that the $\Delta\Phi$ changes very slowly at the beginning and rapidly decreases when $f_\text{isco}$ is approached.
In addition, this phase correction $\Delta\Phi$ is very sensitive to the symmetric mass ratio $\eta$.
Binary systems with very similar components have much greater dephasing $\Delta\Phi$ than those with distinct components.

As discussed in Ref.~\cite{Bonvin:2016qxr}, the $-4$PN correction to the GW phase is very difficult to detect by the ground-based interferometers, such as LIGO/Virgo.
This term is slightly larger for GWs in the LISA band, but, even with a template not considering this effect, the mismatch is less than $10^{-3}$. 
This is due to the extreme smallness of $X(z)$ [refer to Eq.~\eqref{eq-xz}].
Similarly, the correction at 1PN is also highly suppressed by the small Hubble parameters, so it would be even more difficult to detect. 
Finally, the 5PN correction also induces a small mismatch, which will be calculated in the next subsection.

\subsection{Mismatch}

In this subsection, the mismatch due to the 5PN correction is calculated \cite{Apostolatos:1995pj}. 
Although currently the complete waveform at the 5PN has not yet been analytically calculated from the first principle, several phenomenological models exist, which contain terms at and beyond 5PN order, such as IMRPhenonmD \cite{Husa:2015iqa,Khan:2015jqa}. 
In this model, the terms of order higher than 3.5PN in the phase for the inspiral stage were introduced in Eq.~(28) as the ansatz.
Those higher-order terms carry certain phenomenological coefficients, which were fixed by fitting it to the hybrid effective-one-body–numerical-relativity waveforms.
The numerical values for those phenomenological coefficients are further related to the physical quantities, such as the symmetric mass ratio $\eta$ and a spin parameter $\chi_\text{PN}$ [defined in Eq.~(3)], by Eq.~(31) in Ref.~\cite{Khan:2015jqa}, where a new set of coefficients $\lambda_{ij}$ are introduced and tabulated in Table V. 
One can find out that the coefficients $\lambda_{ij}$ for the 5PN term (in the row with $\sigma_3$) have very large absolute values compared to that of the 5PN order correction found in the current work.
So the mismatch caused by the 5PN correction should be small.

In order to calculate the mismatch, the fitting factor $FF$ between two waveforms $h_1$ and $h_2$ is needed. 
Suppose these waveforms are described by a certain number of parameters $\theta^a$, and then $FF$ is defined to be 
\begin{equation}
  \label{eq-def-ff}
  FF=\text{max}_{\Delta\theta^a}\frac{\langle h_1|h_2\rangle}{||h_1||\,||h_2||},
\end{equation}
where $\Delta\theta^a$ represent the differences between the parameters of the waveforms and the numerator on the right-hand side is an inner product
\begin{equation}
  \label{eq-def-ip}
  \langle h_1|h_2\rangle=4\Re\int_0^\infty\frac{\tilde h_1^*(f)\tilde h_2(f)}{S_n(f)}\ud f,
\end{equation}
with $\tilde h_1$ and $\tilde h_2$ the Fourier transformed waveforms and $S_n(f)$ the one-sided noise power spectrum of the interferometer. 
With this inner product, one defines $||h_1||=\sqrt{\langle h_1|h_1\rangle}$ and $||h_2||=\sqrt{\langle h_2|h_2\rangle}$.
The mismatch is thus given by \cite{Apostolatos:1995pj}
\begin{equation}
  \label{eq-def-mm}
  \mathfrak M=1-FF,
\end{equation}
which quantifies the difference between $h_1$ and $h_2$.

In our calculation, we take $\tilde h_1$ to be a simplified IMRPhenonmD waveform whose amplitude is given by 
\begin{equation}
  \label{eq-def-h1-a}
  A\propto \eta^{1/2}(Mf)^{-7/6},
\end{equation}
which is actually the leading-order GW amplitude calculated by considering the quadruple radiation only.
The exact proportionality factor is not important for calculating the mismatch.
The phase of $\tilde h_1$ is given by 
\begin{equation}
  \label{eq-def-h1-f}
  \begin{split}
  \phi_\text{Ins}=&2\pi ft_c-\Phi_c-\frac{\pi}{4}\\
   &+\frac{3}{128}(\pi \mathcal Mf)^{-5/3} \sum_{j=0}^7\varphi_j(\pi Mf)^{j/3}\\ 
   &+\frac{1}{\eta}\left[ \sigma_0+\sigma_1 Mf+\frac{3}{4}\sigma_2(Mf)^{4/3}+\frac{3}{5}\sigma_3(Mf)^{5/3} \right],
  \end{split}
\end{equation}
where the second line includes the fourth term in Eq.~\eqref{eq-gw-ph} and higher PN order corrections up to 3.5PN and the third line includes even higher PN order terms up to 5PN. 
The third line is actually the phase ansatz introduced in Ref.~\cite{Khan:2015jqa}. 
The expressions for the factors $\varphi_j$ and $\sigma_k$ are given in Ref.~\cite{Khan:2015jqa}. 
Although these factors also depend on the spins, we ignore them, i.e., set all spins to zero.
So the parameters describing the waveform $\tilde h_1$ are $M, \eta, t_c$, and $\Phi_c$.
We also ignore the phases for the intermediate and the merger-ringdown stages as discussed in Ref.~\cite{Khan:2015jqa}. 
Finally, the waveform $\tilde h_2$ is basically $\tilde h_1$ modified by the 5PN phase correction given by the last term in Eq.~\eqref{eq-gw-ph}.

We consider the mismatch between the two waveforms observed by ground-based detectors [aLIGO, Virgo, KAGRA, ET \cite{Punturo:2010zza}, and Cosmic Explorer (CE) \cite{Evans:2016mbw}] and the space-borne detector (LISA \cite{Seoane:2013qna}). 
In carrying out the integrations, we set the lower integration limit $f_\text{min}$ for the ground-based detectors to be 1 Hz if Einstein Telescope is used and 5 Hz if other detectors are used. 
The upper integration limit is given by $f_\text{max}=0.018/M$ according to the construction of the inspiral phase in Ref.~\cite{Khan:2015jqa}.
For LISA, the lower integration limit is chosen to be $f_\text{min}=10^{-4}$ Hz, and the upper limit is 
\begin{equation}
  \label{eq-fmax}
  f_\text{max}=\text{min}(f_\text{isco},f_\text{4yr}).
\end{equation}
Here, $f_\text{4yr}$ is the GW frequency evolving from $f_\text{min}$ in 4 years, approximately given by
\begin{equation}
  \label{eq-f4yr}
  f_\text{4yr}=\left[ f_\text{min}^{-8/3}-\frac{256\pi}{5}(\pi\mathcal M)^{5/3}\Delta t \right]^{-3/8},
\end{equation} 
with $\Delta t=4$ years.

\begin{widetext}
 \begin{table}[ht]
  \begin{tabular}{c|cccc|cccccc}
    \hline\hline 
    & \multirow{2}{*}{$m_1(M_\odot)$}& \multirow{2}{*}{$m_2(M_\odot)$} & \multirow{2}{*}{$z$} & \multirow{2}{*}{$d_L$ (Mpc)}  & \multicolumn{6}{|c}{Mismatches}\\\cline{6-11}
    &  &  &  & & aLIGO & Virgo & KAGRA & ET & CE & LISA\\
    \hline
    GW150914 & 35.6 & 30.6 & 0.09 & 424 & $2.48\times10^{-6}$ & $2.29\times10^{-6}$ & $2.02\times10^{-6}$ & $3.76\times10^{-6}$ & $5.90\times10^{-6}$ & -\\ 
    \hline
    GW170817 & 1.46 & 1.27 & 0.01 & 45 & $9.61\times10^{-7}$ & $8.54\times10^{-7}$ & $5.76\times10^{-7}$ & $4.06\times10^{-6}$ & $9.95\times10^{-9}$ & -\\ 
    \hline 
    EMRI & $10^5$ & 10 & 0.20 & 1000 & - & - & - & - & - & $6.05\times10^{-13}$\\ 
    \hline 
    IMRI & $10^5$ & $10^3$ & 0.78 & 5000 & - & - & - & - & - & $1.42\times10^{-8}$\\ 
    \hline 
    IMBH & $5\times10^3$ & $4\times10^3$ & 2.0 & $1.6\times10^4$ & - & - & - & - & - & $3.55\times10^{-15}$\\ 
    \hline
    SMBH & $5\times10^6$ & $4\times10^6$ & 5.0 & $4.8\times 10^4$ & - & - & - & - & - & $7.86\times10^{-8}$\\
    \hline\hline
  \end{tabular}
  \caption{Parameters for different types of GW sources (columns 2 - 5), and the mismatches for different detectors (columns 6 - 11).
  The parameter values for GW150914 and GW170817 are taken from Ref.~\cite{LIGOScientific:2018mvr}, and those in the last four rows are suggested by Ref.~\cite{Chamberlain:2017fjl}.}
  \label{tab-pars}
\end{table}
\end{widetext}

In the calculation, we let the parameters for $\tilde h_1$ be given by the equivalent ones listed in columns 2 to 5 in Table~\ref{tab-pars}.
The values for $t_c$ and $\Phi_c$ play no role in the mismatch, so neither of them is listed. 
The parameters for $\tilde h_2$ differ from those for $\tilde h_1$ by $\Delta\theta^a=(\Delta\mathcal M, \Delta\eta, \Delta t_c, \Delta\Phi_c)$.
Note that $\Delta t_c$ and $\Delta\Phi_c$ need to be specified in the calculation.
Using the MATLAB function \verb+fmincon+ for finding the minimum of a constrained function (since $0<\eta\le0.25$), one can calculate all the mismatches for different detectors.
These values are also tabulated from columns 6 to 11 in Table~\ref{tab-pars}.
One finds out that the mismatches are very small, less than $10^{-5}\sim10^{-6}$. 
Although our 5PN order correction is tiny, it should be considered in the more accurate measurements of GWs.

\section{Conclusion}
\label{sec-con}

In this work, we considered the corrections to the leading-order short-wavelength approximation, by taking into account the temporal and the spatial variation of the GW amplitude. 
The variation is mainly due to the orbital decay of the binary star system; the cosmological evolution also modifies the amplitude, but its impact is much smaller. 
The evolution equations for the leading- and the subleading-order amplitudes were obtained, and these amplitudes were then computed for the GW generated by a binary system. 
The Weyl tensor component $\Psi_4$ was calculated to reveal the modifications to the waveform, in particular, the dephasing.
After the Fourier transformation, the dephasing consists of three contributions. 
The first is at $-4$PN, which was analyzed in Refs.~\cite{Seto:2001qf,Nishizawa:2011eq,Bonvin:2016qxr}.
The  second is at 1PN.
Both of these carry factors containing the source redshift $z$ and the Hubble parameters, but they are too small to be easily detected. 
The final contribution to the dephasing is at 5PN order. 
We quantify the mismatch introduced by this term, and we find that it is typically of the order $10^{-5}-10^{-6}$. 
Although this is tiny, as the accuracy of GW experiments improves, including this term when building templates can be potentially interesting for future generation GW experiments.

\begin{acknowledgements}
  We thank the anonymous referee for improving the clarity of this work. 
This work was supported by the National Natural Science Foundation of China under Grant No.~11633001 and No.~11920101003 and the Strategic Priority Research Program of the Chinese Academy of Sciences, Grant No.~XDB23000000.
\end{acknowledgements}

\bibliographystyle{apsrev4-1}
\bibliography{corrections_GW_amplitudes_v5.bbl}

\end{document}